\documentclass[
reprint,
superscriptaddress,
longbibliography,
groupedaddress, 
amsmath,amssymb,aps,prd,
]{revtex4-2}

\usepackage{graphicx}
\usepackage{dcolumn}
\usepackage{bm}
\usepackage{siunitx} 
\usepackage{booktabs}
\usepackage{multirow} 
\usepackage{tabularx}
\usepackage{comment}
\usepackage[linesnumbered,ruled,vlined]{algorithm2e}
\usepackage{algpseudocode}
\usepackage[mathlines]{lineno}
\usepackage{color,xcolor}
\usepackage{tikz, pgfplots}
\usepackage{listofitems} 
\usetikzlibrary{arrows.meta} 
\usepackage[outline]{contour} 
\contourlength{1.4pt}
\usepackage{hyperref}

\tikzset{>=latex} 
\tikzstyle{node}=[thick,circle,draw=black,minimum size=22,
inner sep=0.5,outer sep=0.6]
\tikzstyle{node in}=[node,black,draw=black,fill=none]
\tikzstyle{node hidden}=[node,black,draw=black,fill=none]
\tikzstyle{node out}=[node,black,draw=black,fill=none]
\tikzstyle{connect}=[thick,black] 
\tikzstyle{connect arrow}=[-{Latex[length=4,width=3.5]},thick,black,
shorten <=0.5,shorten >=1]
\tikzset{ 
  node 1/.style={node in},
  node 2/.style={node hidden},
  node 3/.style={node out},
}

\newcommand{\MIT}{Massachusetts Institute of Technology, Cambridge, Massachusetts 02139, USA}

\makeatletter
\def\maketitle{
\@author@finish
\title@column\titleblock@produce
\suppressfloats[t]}
\makeatother

\begin{document}

\preprint{APS/123-QED}

\title{Discovering Nuclear Models from Symbolic Machine Learning}

\author{Jose M. Munoz}
\email{josemm@mit.edu}
\author{Silviu M. Udrescu}
\email{sudrescu@mit.edu}
\author{Ronald F. Garcia Ruiz}
\email{rgarciar@mit.edu}
\affiliation{\MIT}


\begin{abstract}

Numerous phenomenological nuclear models have been proposed to describe specific observables within different regions of the nuclear chart. However, developing a unified model that describes the complex behavior of all nuclei remains an open challenge. Here, we explore whether novel symbolic Machine Learning (ML) can rediscover traditional nuclear physics models or identify alternatives with improved simplicity, fidelity, and predictive power. To address this challenge, we developed a Multi-objective Iterated Symbolic Regression approach that handles symbolic regressions over multiple target observables, accounts for experimental uncertainties and is robust against high-dimensional problems. As a proof of principle, we applied this method to describe the nuclear binding energies and charge radii of light and medium mass nuclei. Our approach identified simple analytical relationships based on the number of protons and neutrons, providing interpretable models with precision comparable to state-of-the-art nuclear models. Additionally, we integrated this ML-discovered model with an existing complementary model to estimate the limits of nuclear stability. These results highlight the potential of symbolic ML to develop accurate nuclear models and guide our description of complex many-body problems.

\end{abstract}
\maketitle
\section*{Introduction}
\label{sec:Introduction}
The atomic nucleus, a strongly correlated quantum many-body system, exhibits a vast array of phenomena that cannot be explained by a single, unified theory. This is mainly because the underlying theory of the nuclear force, quantum chromodynamics (QCD), is non-perturbative at low energies~\cite{tews2020new}. Despite the exciting progress that has been made in developing a comprehensive characterization of the nuclear force and powerful many-body methods, a theoretical description of all the nuclei, with a direct link to QCD, remains a major unsolved challenge \cite{RevModPhys.81.1773}.
Consequently, physicists have developed a myriad of models, each designed to address specific aspects of nuclear structure and reactions. These models range from the simple liquid drop model to sophisticated many-body calculations, which are continuously refined to reproduce available nuclear data, such as binding energies, nuclear radii, electromagnetic moments, excitation energies, and decay properties \cite{Nav16,Rei17,Sassarini:2021bxm,Tews:2020hgp, lovato2022hidden,Ver22,Kar23}. Thus, one hopes that improved models can provide an interpretation of the observed phenomena and predict the behavior of nuclei at the extremes of the nuclear chart, where data is not yet available.

Recent advancements in Machine Learning (ML) have emerged as a pragmatic tool to tackle this challenge, with the expectation of fitting models that could reproduce the vast amount of available nuclear properties~\cite{gao2021machine, su2023progress, neufcourt2018bayesian, Akkoyun:2021mce,mumpower2023bayesian,lovato2022hidden,Neufcourt:2018syo}. Nonetheless, this approach results in black boxes with obscure physical interpretability~\cite{Munoz:2022slm, nolte2023nuclr, PhysRevC.106.L021301}, and often its use is limited to specialized users who can have access to and run specific software. Moreover, conventional ML models rely on extensive experimental data or high-fidelity emulators, which may not always be available and are very prone to over-fitting, drastically limiting their extrapolation power~\cite{belkin2018overfitting, PhysRevC.100.054326}.

An alternative approach is given by symbolic ML, which aims to provide analytical expressions to describe a given observable~\cite{gerwin1974information}. Intriguingly, this methodology is remarkably similar to how nuclear physicists have traditionally tackled the problem of describing complex nuclear phenomena. Unlike traditional regression techniques that fit data to a pre-specified model, such as linear or polynomial functions, modern symbolic regression approaches do not assume any particular prior model. In contrast, it changes the existing functional forms in an evolutive manner~\cite{doi:10.1126/science.1165893, disc_keren, keijzer2004scaled}. Because of this, discovering complex equations is now conceivable, something that could not be accomplished by evaluating every possible mathematical expression by brute force~\cite{doiudrescu2020ai,udrescu2020ai}. These methods have been widely successfully applied in scientific research, including Cosmology~\cite{CAMELS:2020cof, tenachi2023deep}, Astronomy~\cite{matchev2022analytical, lemos2023rediscovering}, Materials science~\cite{wang2019symbolic}, Chemistry~\cite{sr_chem}, Dynamics~\cite{DERNER2020106432}, and Particle physics~\cite{dong2023machine, Tsoi:2024ypg}. However, applying existing methods to describe nuclear physics observables poses several challenges, as most of them have not been developed to solve multi-objective problems and/or do not consider experimental uncertainties.

In this work, we explore the natural question of whether symbolic ML approaches, with minimal human bias, can rediscover traditional nuclear physics models or propose alternative ones with similar or better simplicity, fidelity, and predictive power.
To achieve these goals, we developed a Symbolic Regression technique that can systematically handle regressions over a set of targets, considers experimental uncertainties, and is robust to high-dimensional problems.

Our approach achieves three key results. First, we provide an interpretable model that discovers relatively simple analytical relationships between the number of nucleons and nuclear properties such as the charge radius and binding energy. Second, we use ML techniques to estimate these nuclear observables beyond our current experimental knowledge. Finally, an improved model, with quantifiable uncertainties, is used to provide predictions of the limits of stability of both neutron-deficient and neutron-rich nuclei.

\section*{Symbolic Regression for Nuclear Observables}
\begin{figure}[!h]
\centering
\includegraphics[width=0.5\textwidth]{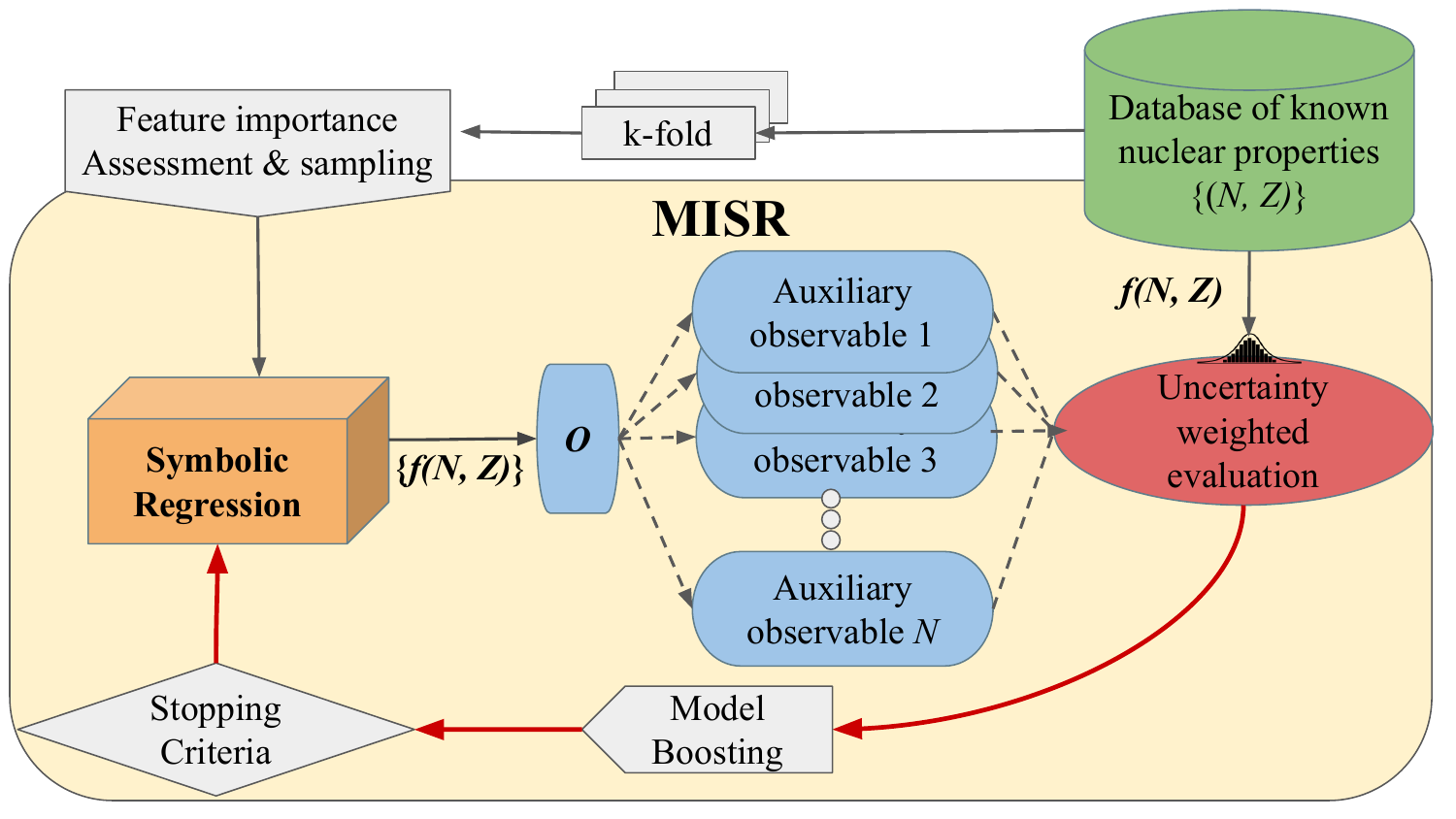}
    \caption{Diagrammatic representation of Multi-objective Iterated Symbolic Regression (MISR) inner pipeline. The process iteratively refines the symbolic regressed models. See the text for more details.}
    \label{fig:MISR}
\end{figure}

We developed a Multi-objective Iterated Symbolic Regression (MISR) to iteratively search for analytical models that best describe a set of target variables up to arbitrary accuracy, leading to an expansion-like expression. We build on top of symbolic regression methods~\cite{cranmer2020discovering} to adopt multi-objective capabilities, which allows us to fit the description of a set of diverse target variables, considering the associated experimental uncertainty for each variable. A diagrammatic representation of the MISR approach is shown in Fig.~\ref{fig:MISR}. It is worth noting the MISR is in principle agnostic to the SR tool to be employed, as it only constitutes a block of the pipeline.
As illustrated in the figure, the main variable, regressed via the standard symbolic regression, is then used to trace over the Pareto-optimal expressions for minimizing the error over the auxiliary variables in the several $k$ random sub-samples of the data (known as $k$-folds). This process is then repeated using boosting to fit the residuals of the current analytical model until we reach a stopping criteria~\cite{schapire2003boosting, ngatchou2005pareto}. Part of these advances were motivated by the application of multi-objective approaches for dynamic modeling~\cite{kubalik2021multi}. 

Unlike traditional regression techniques that offer a static view of feature relevance, MISR continuously re-evaluates in each boosting step the feature importance via both Boosted Decision trees and Mutual Information Regression~\cite{schapire2003boosting, kozachenko1987sample}. This allows each one of the obtained terms to represent a particular behavior within the modeled phenomena. This process not only enhances the predictive power but also streamlines the model by eliminating redundant or less significant features.

One of the critical challenges in ML approaches is the quantification of uncertainty in predictive modeling, especially given the often limited and low-precision experimental data that exists for certain nuclear properties. To achieve an estimation of uncertainty for the MISR model, we resample different training datasets using jackknife bootstrapping~\cite{alaa2020discriminative} to link each one of the model's predictions to a probability distribution, from which we sample the posteriors. In addition, we include a truncation uncertainty, for a specific cutoff, given by the absolute value of the subsequent expansion term. This methodology was inspired by an analogy to how effective theories are developed~\cite{furnstahl2015quantifying}.
 
This iterative refinement allows us to express the target data as a sum of the resultant $n$ analytical forms $f^{(i)}(N, Z)$, and estimate the uncertainty of the model. The MISR can thus be expressed as
\begin{equation}
    \text{MISR}n \ (N, Z) = \sum_{i=1}^n \mathcal{N}(1, \sigma_i)\cdot f^{(i)}(N, Z),
\end{equation}
where $N$ and $Z$ are the input variables, representing the number of neutrons and protons in the nuclei, respectively. 
$\mathcal{N}(1, \sigma_i)$ is a Gaussian distribution numerically fitted after the training process using the discriminative jackknife technique \cite{alaa2020discriminative}.

We define the nuclear mass as $A=N+Z$, isospin asymmetry, $I= (N - Z)/A$, and the Casten factor 
$P= (N_n N_p)/(N_n+N_p)$, where $N_p$ and $N_n$ represent the valence protons and neutrons filling pre-defined shell-model orbits, i.e. the difference between the number of nucleons and the closest magic number $\{8,20,28,50,82\}$ \cite{PhysRevLett.58.658}. The model was trained using the input variables $\{(N, Z, A, I, P, N_n, N_p)\}$. More details of the methodology can be found in the section Methods~\ref{sec:SI-MISR}.

As a proof-of-principle of our approach, we focus on the study of nuclear binding energies and charge radii. However, our proposed method can be generally applied to describe other nuclear observables, with an arbitrary set of input variables. 
Binding energy, the energy required to disassemble a nucleus into its constituent protons and neutrons, provides important information to guide our understanding of atomic nuclei. This observable is arguably one of the most studied nuclear observables, with a relatively large collection of experimental data available.
To train our algorithm, we used a subset of the experimental data of binding energies reported in the AME2020 mass evaluation dataset~\cite{wang2021ame, 2023APS..HAWM06002B, DFTMassTables}, and experimental charge radii reported in~\cite{MarinovaAngeli2013}. 
For the charge radii —characterizing the charge distributions inside the nuclei— there is a more scarce set of experimental measurements~\cite{yang23}. This is particularly challenging for light isotopes, which, combined with the large variations in their radii, motivated us to focus on nuclei with $12 \leq Z \leq 50$, for which reliable experimental data exists.
From these data sets, we sampled uniformly $20\%$ of the nuclei for testing purposes. The cutoff on light and medium mass nuclei allows us to study the extrapolation capabilities of our algorithm and provide a test of the model's ability to deduce complex nuclear properties from relatively small data sets. This demonstrates the broader applicability of our MIRS method, which can describe physical observables where data scarcity is common.

\subsection*{Results for nuclear binding energies}

The study of nuclear binding energies has led to the proposal of several analytical formulas. One of the simplest, the Liquid Drop Model (LDM) \cite{benzaid2020bethe}, represents the nucleus as a liquid drop to explain its global properties as a function of $Z$ and $A$:
\begin{equation}
    BE_{LDM} =\alpha_v A -\alpha_s A^{2/3} -\alpha_c \frac{Z^2}{A^{1/3}} -\alpha_a \frac{(A - 2Z)^2}{A} + \delta,
\end{equation}
where \( a_v \), \( a_s \), \( a_c \), and \( a_a \) are the volume, surface, Coulomb, and asymmetry coefficients, respectively, and \( \delta(A,Z) \) represents the pairing term between neutrons and proton computed as \begin{equation}
\delta(A,Z) =
\begin{cases}
+\delta_0 A^{-1/2} & \text{if Z and N are even}, \\
0 & \text{if Z or N is odd }, \\
-\delta_0 A^{-1/2} & \text{if Z and N are odd},
\end{cases}
\end{equation} where \( \delta_0 \) is another fitted constant~\cite{benzaid2020bethe}. 
More complex empirical models such as the Finite Range Droplet Model (FRDM) \cite{moller2012new}, and the Duflo-Zuker (DZ) \cite{Mendoza-Temis:2009uje}, attempt to incorporate microscopic corrections to these macroscopic trends, refining the reproduction of nuclear binding energies across the chart of nuclides. The DZ model integrates both macroscopic and microscopic aspects of nuclear structure, featuring a wide array of terms to capture various models of nuclear interactions and nucleon configurations. The DZ model is not represented by a concise formula, but rather as a computational tool. This model is relatively complex and includes several terms to numerically account for nuclear shell effects, monopole Hamiltonian contributions, deformation energies, symmetry, and Coulomb energies.

State-of-the-art nuclear models, such as Density Functional Theory (DFT) \cite{Erl12} and ab initio calculations \cite{PhysRevLett.126.022501}, follow a very different approach. These models aim to provide a microscopic description of nuclei, starting from inter-nucleon interactions and solving numerically the nuclear quantum many-body problem. DFT, in particular, has been very successful in describing the binding energy and radii across the nuclear chart \cite{Rei17,Perera:2021ztx}. For ab initio methods, the simultaneous reproduction of both binding energy and nuclear charge radii remains an open challenge \cite{Tews:2020hgp,Mal22}.

To explore the multi-objective capabilities of our model, we apply the MISR algorithm to predict the binding energy, using as auxiliary observables the binding energy per nucleon, and the one and two nucleon separation energies across the same $Z$ region, defined as
$S_n=BE(Z, N+1) - BE(Z, N)$, $S_{2n}=BE(Z, N+2) - BE(Z,N)$, $S_p=BE(Z+1, N) - BE(Z,N)$, $S_{2p}=BE(Z+2, N) - BE(Z,N)$.

At leading order, in the first iteration of MISR, we obtained a relatively simple expression: 
\begin{equation}\label{eq:eb_first}
    \text{BE}_{MISR}^{(1)} =\eta_0  Z\left(1+\frac{1}{N}-\frac{a N}{Z^2}\right) \left[I \left(b\, -\frac{{A^{1/3}} N}{Z}\right)+c\right]
\end{equation}
with  $a=-1.10$, $b=32.43$, $c=16.70$, and $\eta_0=1.0$ MeV. Additional MISR iterations, up to $n=10$, are presented in the section Methods, Appendix~\ref{tab:be_misr_eqs}.

\begin{figure}[!h]
     \centering
     \includegraphics[width=0.5\textwidth]{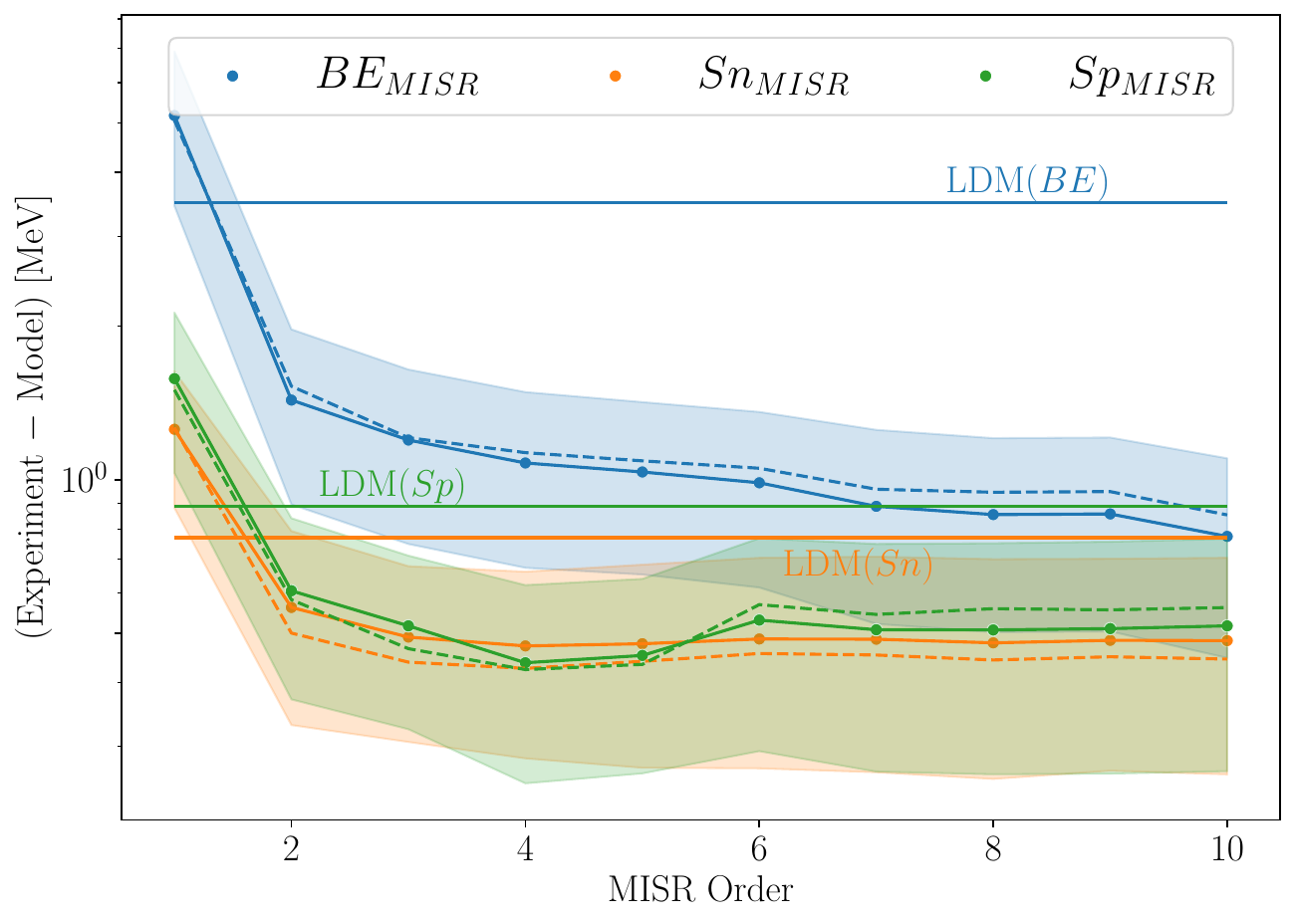}
     \caption{Convergence of different observables fitting the MISR on the nuclear binding energies. The colored area illustrates the standard deviation of the residuals among the training set and the dashed line shows the mean over the test nuclei. The LDM results are illustrated as horizontal lines for reference.}
     \label{fig:BE_SR_errors}
 \end{figure}

The differences between the experimental values and model predictions for various observables are shown in Fig.~\ref{fig:BE_SR_errors}. These differences decrease as one includes more orders of the MISR expansion, illustrating how the regression converges without a sign of overfitting to the trained region. In principle, one can keep an arbitrary number of terms in the expansion. However, the Pareto Frontier hints it is optimal to keep up to 10 terms.

As illustrated in Fig.~\ref{fig:BE_SR_errors}, our MISR results for $n=2$ exhibit a significant improvement in the prediction of nuclear binding energies with respect to the LDM. This is also true for the neutron and proton separation energies.

\begin{figure}[!h]
\centering
\includegraphics[width=0.5\textwidth]{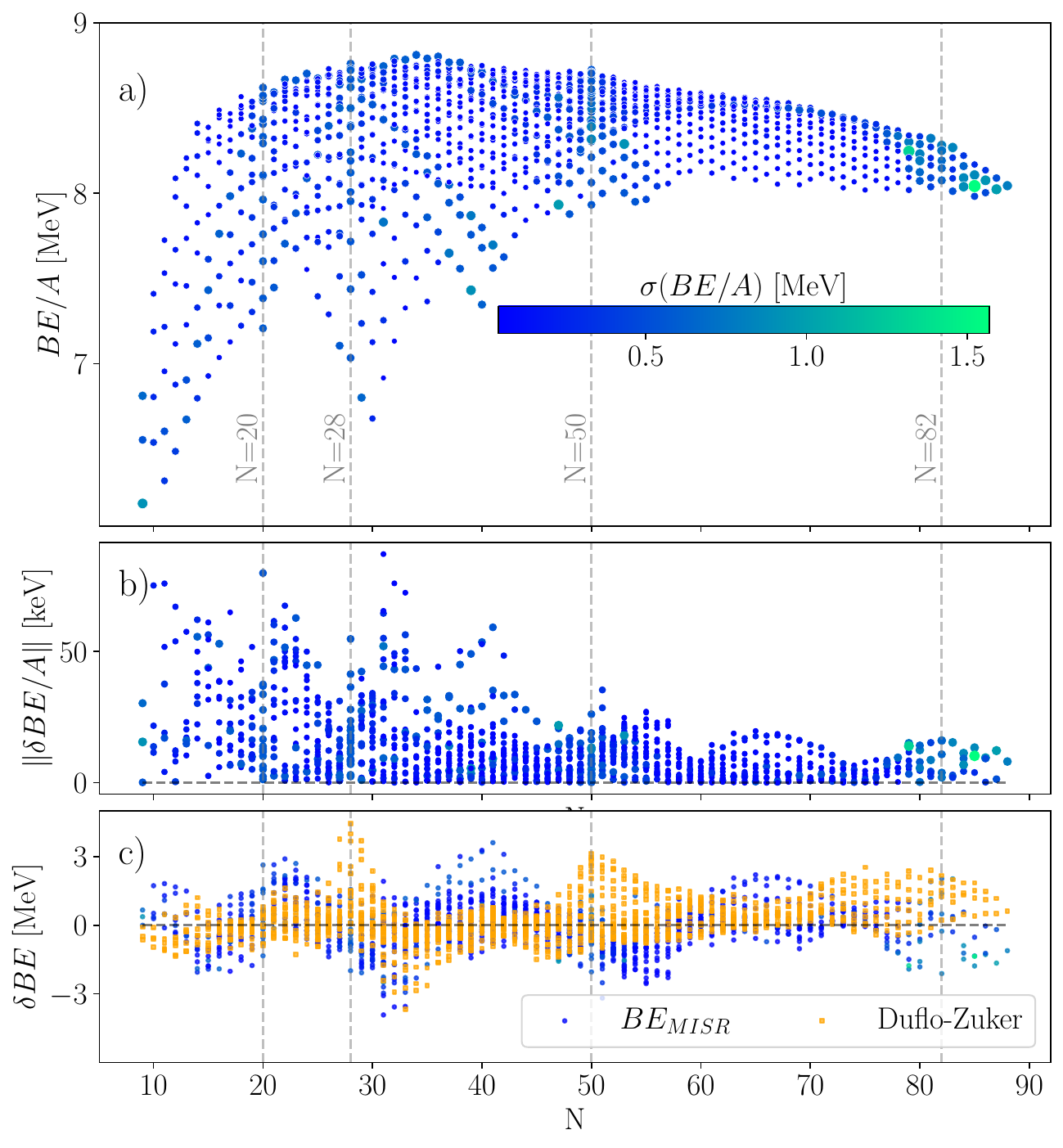}
\caption{a) Predicted binding energy per nucleon $(BE/A)$ as a function of neutron number with associated model uncertainties denoted by the color bar. b) Absolute error on $\|BE/A\|$ predictions (Experiment - $BE_{MISR}$), showcasing the distribution of discrepancies for different neutron numbers. c) Residuals of the binding energy (Experiment - Model)  obtained for our MISR10 results and the Duflo-Zucker models. Vertical dashed lines show the traditional nuclear magic numbers.}
\label{fig:BE_A}
\end{figure}

The results for the binding energy per nucleon, $BE/A$, together with the distribution of their predicted uncertainties are shown in Fig.~\ref{fig:BE_A}. The ability to provide uncertainty estimates allows for a more robust analysis of the predictive power of the model. This feature of MISR assesses where additional experimental data or refinement of the model might be necessary. The standard deviation \( \sigma(BE) \) of the binding energy predictions, as indicated by the color gradient, tends to correlate with the residual distribution. We remark that the uncertainty is not uniform across the nuclear chart. Instead, the highest uncertainties coincide with the transitional regions between different nuclear shells. These regions are known for increased structural changes that impact nuclear stability and are more challenging to predict accurately. This shell structure effect is illustrated in Fig.~\ref{fig:BE_A} c). The MISR expansion is well-behaved around closed-shell structures but its accuracy decreases for open-shell nuclei. The contrary is true for the Duflo-Zuker model, with the discrepancy with respect to the experiment larger around neutron magic numbers. 

In Table \ref{tab:mae_models_be} we compare the performance of different nuclear models in terms of Mean Absolute Error (MAE) and Root Mean Squared Error (RMSE). Note that these results represent only a subset of DFT calculations and do not include the latest DFT developments. MAE is calculated as the average absolute difference between the predicted and experimental values, while RMSE is derived from the square root of the average of the squared differences between the predicted and experimental binding energy values. For comparison, the table includes the results from DFT calculations using different density functionals such as DD-ME2 \cite{PhysRevLett.126.032502}, NL3* \cite{PhysRevC.89.054320}, HFB24 \cite{goriely2016latest}, and UNEDF1 \cite{Kortelainen_2012}. Remarkably, the MISR10 model shows competitive performance, particularly when considering the fact that it is built on a small number of input variables. 
\begin{table}[ht]
    \centering
    \caption{Mean Absolute Error (MAE) and Root Mean Squared Error (RMSE) obtained for the $BE$ of nuclei with $12 \leq Z \leq 50$. The results from MISR1 and MISR10 are compared with different nuclear models. See text for more details. }
    \label{tab:mae_models_be}
    \begin{tabular}{lc|cr}
        \hline\hline
         Model&Ref. & MAE [MeV] & RMSE [MeV] \\
        \hline
        {MISR1} &-&5.11	&6.17\\
        {MISR10} &-& 0.78 & 0.99 \\
                DD-ME2 & \cite{PhysRevLett.126.032502}& 2.48 & 2.83 \\
        NL3* & \cite{PhysRevC.89.054320}& 1.95 & 2.45 \\
        Duflo-Zuker & \cite{Mendoza-Temis:2009uje}&0.62 & 0.87 \\
        FRDM & \cite{moller2012new} &0.67 & 0.90 \\
        \textbf{HFB24} & \cite{goriely2016latest}& 0.55 & 0.73 \\
        UNEDF1 & \cite{Kortelainen_2012}& 1.88 & 2.25 \\
        LDM & \cite{benzaid2020bethe}&3.56 & 5.08 \\
        \hline\hline
    \end{tabular}
\end{table}

\subsection*{Results for the nuclear charge radii}
Now we shift our focus to the study of the nuclear charge radius, $r_c$. This nuclear property is highly sensitive to the details of the inter-nucleon interactions, not yet fully understood. The reproduction of the magnitude of $r_c$ is an unresolved challenge for ab initio nuclear theory~\cite{rossi2021progress}.


At leading order, we obtain the first MISR expression to be:
\begin{gather}\label{eq:rc_first}
    r_{MISR}^{(1)} =  A^{1/3} \left(a + \frac{b}{Z} \right) + c\left(  P - \frac{Z}{N}\right) e^I- I,
\end{gather}
with $a=0.95$ fm, $b=1.48$ fm, $c=0.017$ fm. Additional terms obtained for higher order corrections, up to $n=10$, can be found in the section Methods, Appendix~\ref{tab:rc_misr_eqs}.
The first term is similar to the well-known droplet model for the isospin symmetric case, $r_c\sim A^{1/3}$. Interestingly, the second term in this equation shows an exponential dependence on the isospin asymmetry, $I$, which, to our knowledge, does not appear in other phenomenological models for the charge radii. Similar exponential terms have been proposed in nuclear mass models~\cite{SEEGER19611}. 

\begin{figure}[!h]
    \centering
    \includegraphics[width=0.5\textwidth]{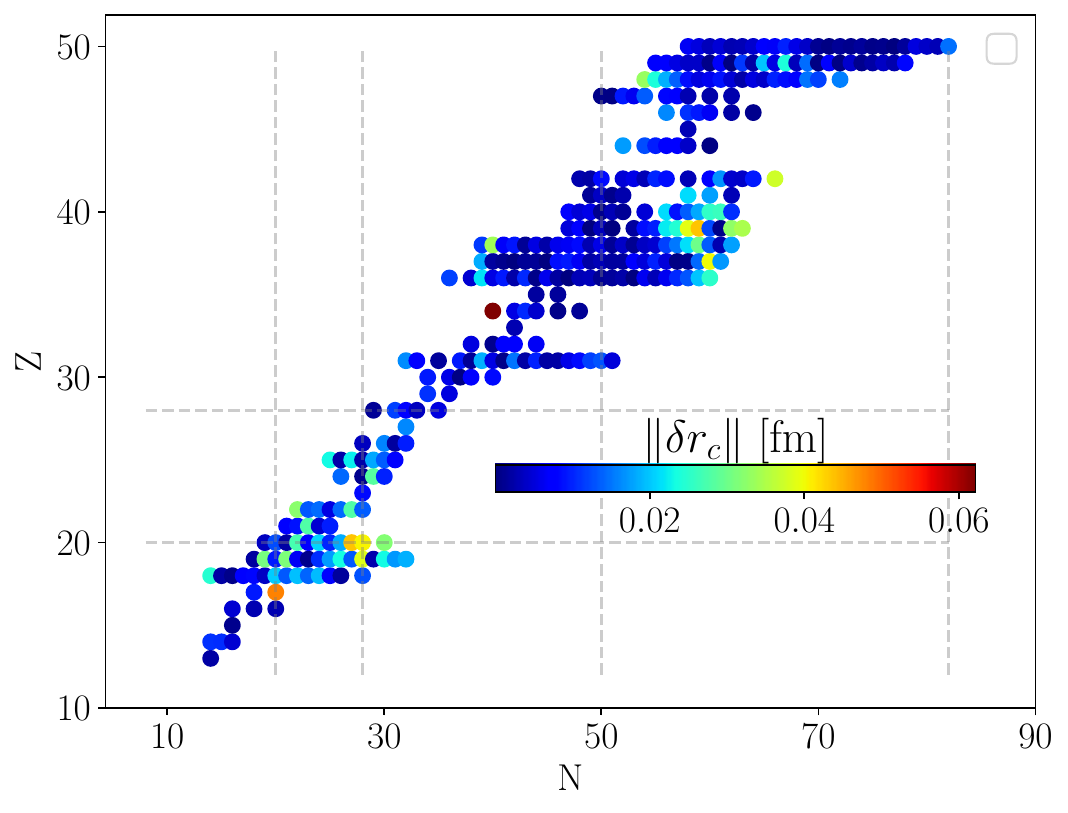}
    \caption{Charge radii differences between the experimental value and prediction obtained by the MISR10 model. The magnitude of these differences is shown with different colors as a function of the neutron and proton numbers.}
    \label{fig:dist_error}
\end{figure}

The difference between the experiment and our MISR results is shown in Fig.~\ref{fig:dist_error}. The MISR expansion captures the overall trend of the charge radii. We do find a higher residual outlier in the distribution for $Z=34, N=40$ ($^{74}$Se), which is mainly due to its relatively large experimental uncertainty. Similarly, as with the results obtained for binding energies, the residuals are larger between magic numbers.

The performance obtained on the training and testing sets are shown in Table \ref{tab:mae_models_rc}. Our MISR results are contrasted with different DFT calculations in the same table. For comparison, we include an analytical model proposed by Nerlo-Pomorska with Casten modifications~\cite{Sheng2015}:
\begin{equation}
    r_c^{NP} = r_0 A^{1/3}\left[1 - \alpha_1\frac{N - Z}{A} + \left(\frac{\alpha_2}{A}\right)^{1/3} + \frac{\alpha_3 P}{A} +\frac{\alpha_4 \delta}{A}\right],
\end{equation}
where the constants $\alpha_i$ are fitted to the experimental data. 

\begin{table}[ht]
    \centering
    \caption{Mean Absolute Error (MAE) between experiment and model for the charge radii of nuclei with $12 \leq Z \leq 50$. We label MISR1 as the first iteration's result of the MISR. Our results are compared with different DFT calculations and the NP formula. See text for more details.}
    \label{tab:mae_models_rc}
    \begin{tabular}{lc|cr}
        \hline\hline
         Model&Ref. & Train [fm] & Test[fm] \\
        \hline
          \textbf{MISR1} &-& \textbf{0.017} & \textbf{0.014} \\
        \textbf{MISR10} &-& \textbf{0.009} & \textbf{0.009} \\         
        DD-ME2 & \cite{PhysRevLett.126.032502}& 0.019 & 0.016 \\
        NL3* & \cite{PhysRevC.89.054320}& 0.028 & 0.026 \\
        SKMS & \cite{buskirk2023nucleonic}& 0.019 & 0.016 \\
        UNEDF1 & \cite{Kortelainen_2012}& 0.026 & 0.020 \\
         NP & \cite{Sheng2015}&0.023 & 0.018 \\
        \hline\hline
    \end{tabular}
\end{table}

Our MISR for the nuclear charge radii exhibits an overall good agreement with the experimental data, which is better than all models presented in Table \ref{tab:mae_models_rc}. This is the case even at leading order, Eq.~\ref{eq:rc_first}, with an expression of remarkable simplicity.

\begin{figure}[!h]
    \centering
    \includegraphics[width=0.5\textwidth]{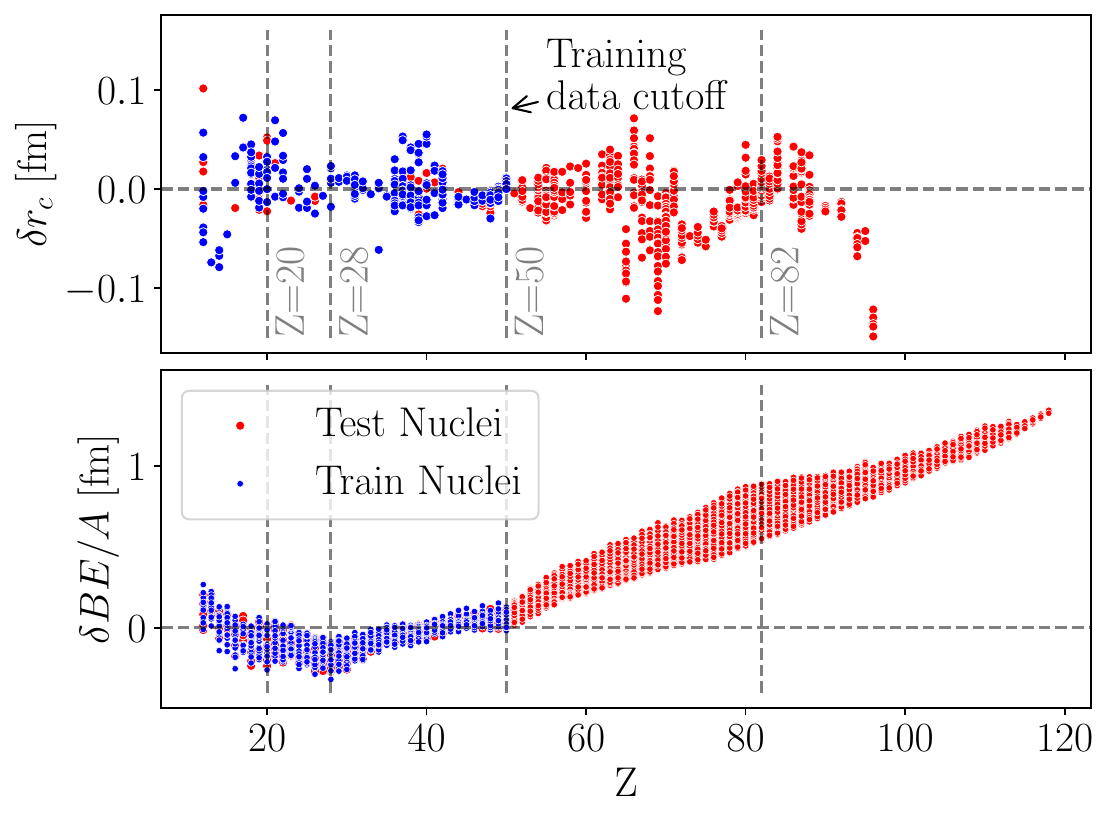}
    \caption{Top: Distribution of the residual on the charge radii as a function of $Z$ for the first term in MISR (Experiment - $r_{MISR1}$). Bottom: Same but for binding energy (Experiment - $BE_{MISR1}$). MISR was trained only using the blue points.}
    \label{fig:rc_extrapolation}
\end{figure}

While Equation~\ref{eq:rc_first} is obtained by training only on light nuclei, it extrapolates well to heavy nuclei. This is illustrated in Fig. \ref{fig:rc_extrapolation}. This implies that our multi-objective optimization yields physical information valid across the entire nuclear chart. A similar extrapolation is observed for the binding energy predicted by Equation~\ref{eq:eb_first}, which shows a smooth linear increase with the proton number. This suggests that adding a simple linear term in $Z$ can provide a major improvement for heavy nuclei. Such a term would likely appear if large nuclei were included in the training process. We choose not to add this term manually, as our goal is to allow the ML algorithm to discover analytical expressions with minimal human intervention.

\begin{figure}[!h]
    \centering
    \includegraphics[width=.5\textwidth]{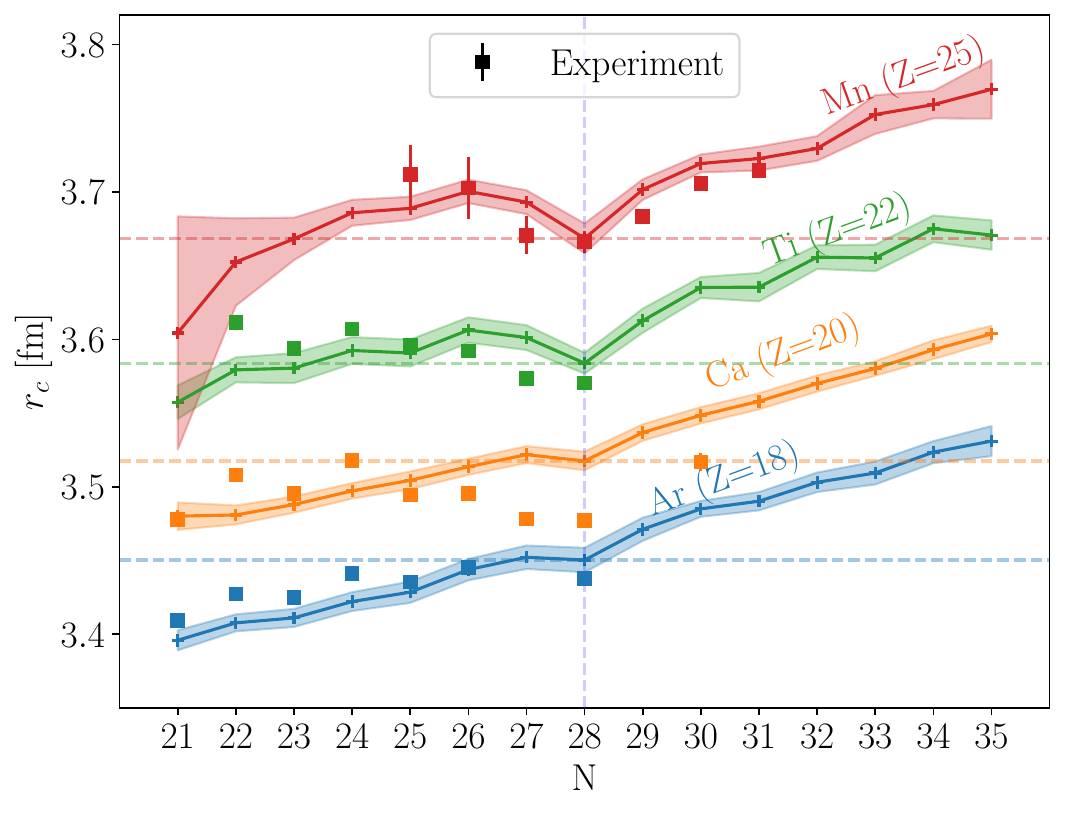}
    \caption{Charge radii trends of elements with $Z=18, 20, 22, 25$. Experimental results \cite{ANGELI201369,garcia2016unexpectedly,Koszor_s_2021,Hey16} are contrasted with our MISR10 results. }
    \label{fig:range_z_19}
\end{figure}

In contrast to nuclear binding energies, the nuclear charge radii are known to exhibit more complex and distinct behavior across the nuclear chart. In what follows, we explore the performance of our MISR result in describing complex charge radii trends for selected isotopic chains with $Z=18, 20, 22, 25$. Our results are compared with the experimental values in Fig. \ref{fig:range_z_19}. 

Overall, the MISR prediction, including uncertainty estimation, provides a good description of the experimental trends. The model captures the expected kinks at magic numbers and part of the odd-even staggering effects, with large discrepancies found for Ca ($Z=20$). These isotopes exhibit a unique trend, which is known to be challenging to describe by nuclear theory \cite{garcia2016unexpectedly,Koszor_s_2021}.

\begin{figure}[!h]
    \centering
    \includegraphics[width=.5\textwidth]{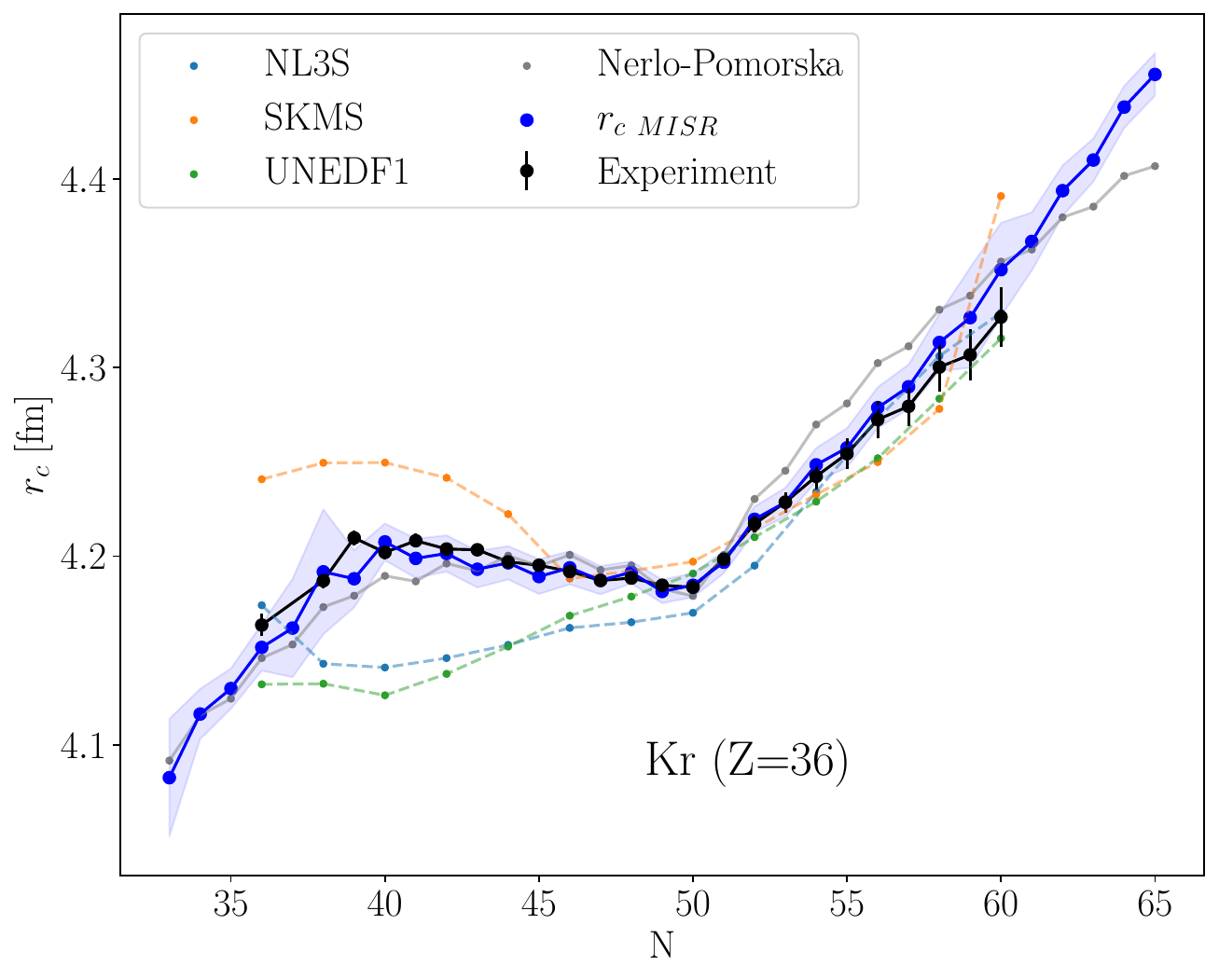}
    \caption{Nuclear charge radii values for the Krypton isotopic chain. Experimental results \cite{ANGELI201369} are compared with different nuclear models. The colored area represents the uncertainty of the MISR model.}
    \label{fig:range_z_36}
\end{figure}

A different, distinct charge radii trend is observed for Krypton (Kr) isotopes ($Z=36$). The MISR results are compared with the experiment in Fig.~\ref{fig:range_z_36}. Here, the MISR model performs remarkably well, capturing all the main features: i. the kink at $N=50$; ii. a parabolic trend for neutron-deficient isotopes; iii. staggering between isotopes with odd and even neutron numbers.

As illustrated in Fig. \ref{fig:range_z_36}, the analytical formula (NP) from \cite{Sheng2015} also performs relatively well for the particular case of Kr isotopes. The difference between the NP and MISR becomes significant for neutron-rich nuclei.


\subsection*{Limits of the nuclear landscape}

Inspired by the fact that our MISR model appears to be highly complementary to the DZ model, i.e. MISR exhibits better performance around closed shells, while the DZ performs better for open-shell nuclei (see Fig. \ref{fig:BE_A}), we combined them as an ensemble model. Both models can be combined using a Bayesian Automatic Relevance Determination (ARD) regression~\cite{mackay1994bayesian}, which can determine the weight of each model and provide a combined prediction with uncertainty. The obtained assembled model, labeled as ARD provides an MAE of 0.389 (0.411) MeV on the train (test) samples, yielding a relative improvement of around $20\%$ with respect to the best-performing HFB24 model \cite{goriely2016latest}, a comparison of different models is presented in Table~\ref{tab:err_ard}.

\begin{table}[ht]
    \centering
    \caption{Mean Absolute Error (MAE) between experiment and model for binding energy and separation energies of nuclei with $12 \leq Z \leq 50$ over the complete set of nuclei (testing and training). Our results are compared with different DFT calculations and phenomenological models. See text for more details.}
    \label{tab:err_ard}
    \begin{tabular}{lc|c|c|c}
        \hline\hline
         Model&Ref. & $BE$ [MeV] & $S_n$[MeV] & $S_p$[MeV] \\
        \hline
        {MISR10} &-& 0.79 & 0.46 & 0.96\\
        \textbf{ARD} &-& 0.39 & 0.28 & 0.77 \\
        Duflo-Zuker & \cite{Mendoza-Temis:2009uje}&0.62 & 0.30 & 0.81 \\
        FRDM & \cite{moller2012new} &0.67 & 0.34 & 0.84\\
        HFB24 & \cite{goriely2016latest}& 0.55 & 0.48 & 0.91 \\
        UNEDF1 & \cite{Kortelainen_2012}& 1.88 & 0.41 & 0.82 \\
        \hline\hline
    \end{tabular}
\end{table}

\begin{figure*}[!ht]
    \centering
    \includegraphics[width=0.8\textwidth]{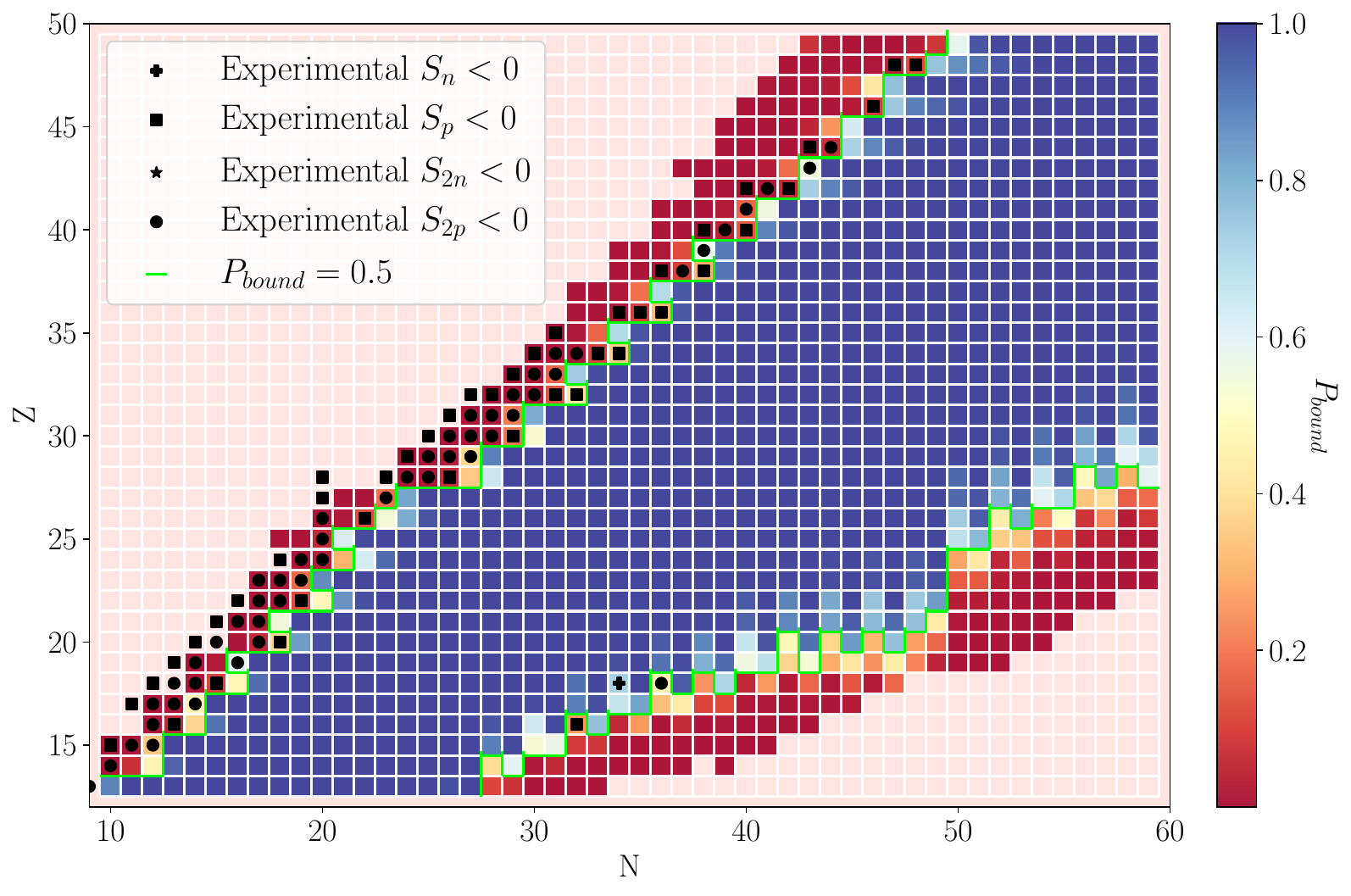}
    \caption{\centering Predict limits of stability obtained with the ARD model. The colors represent the estimated probability of a bound nucleus for a particular combination of $N$ and $Z$. The green lines represent the threshold of having a bound nucleus, defined as $P_{bound}\geq 0.5$. Experimental limits are shown with black symbols.}
    \label{fig:drips}
\end{figure*}

This motivates us to use the ARD model to study open questions in nuclear science, such as the limits of the existence of nuclear matter. The ARD model provides a robust estimation of nucleon separation energies, which is critical for predicting the limits of stability of nuclei. Using our model uncertainties, the probability of having a bound nucleus can be estimated for any number of protons and neutrons. Proceeding with this approach, we compute the one and two nucleon separation energies and obtain the probability of having a positive central value via the Cumulative Density Probability of a Gaussian distribution~\cite{PhysRevLett.126.022501}. The results for the limits of stability predicted by the ARD model for nuclei with $12 \leq Z \leq 50$ are presented in Fig. \ref{fig:drips}. Where data are available, overall good agreement with the experiment was found.

\section*{Conclusions and outlooks}

We developed a Multi-objective Iterative Symbolic Regression framework, labeled as MISR, to enable the discovery of analytical models of nuclear properties. As a proof of principle, the MISR method was employed to describe the binding energies and charge radii of light and medium mass nuclei. Remarkably, simple expressions were found as functions of the neutron and proton numbers. The models found can provide a relatively good description of the available data, with precision comparable to that of state-of-the-art nuclear models.

The MISR model was combined with the well-known DZ model to enable a powerful, complementary description of binding energies and nucleon separation energies. The combined model, ARD, resulted in an overall agreement with the experiment. The uncertainty estimation provided by the ARD model was used to estimate the limits of stability of nuclei in the region $12 \leq Z \leq 50$. These results highlight the potential of integrating physics-informed ML approaches with established complementary theoretical models to improve predictions of yet-to-be-explored regions of the nuclear chart.

By combining the predictive power of machine learning with the interpretability of symbolic expressions, MISR offers a promising avenue for advancing our understanding of complex nuclear phenomena and, hopefully, guiding the development of more accurate and insightful nuclear models.

Future work will focus on extending the application of MISR to describe complementary nuclear observables using different sets of input variables. Of particular interest would be the development of the MISR method to discover the analytical forms of nuclear interactions and, more generally, interaction Hamiltonians of quantum many-body systems. This could be combined with powerful many-body methods to establish a direct link between microscopic interactions and observables. Work is ongoing to generalize the MISR approach for this purpose, including vectors and tensor operators.

\section*{Acknowledgements}

This work was supported by the Office of Nuclear Physics, U.S. Department of Energy, under grants DE-SC0021176 and DE-SC0021179. We are grateful for useful discussions and suggestions from S. Wilkins, J. Holt, A. Belley, A. Galindo, and K. Matchev for their helpful comments and insights.

\bibliography{main}
\clearpage

\section{METHODS}
\subsection{Multi-objective Iterated Symbolic Regression}
\label{sec:SI-MISR}

Symbolic Regression (SR) methods optimize both the parameters and structure of analytical models and have been increasingly applied across various fields such as physics, biology, and finance due to their interpretability and ability to produce simple mathematical expressions \cite{cranmer2020discovering}. SR evolves a population of candidate solutions by crossover and elimination, based on fitness criteria to better fit the data. Moreover, it differs from traditional regression methods in that it does not assume a predefined model structure. Instead, it evolves the structure of the model itself, allowing for discoveries of novel relationships between variables, while requiring significantly less data to train \cite{wilstrup2021symbolic}. In the following, we present Multi-objective Iterated Symbolic Regression  (MISR), a novel framework for performing SR.

\subsubsection*{Overview of the Core Algorithm}
The core algorithm of MISR is rooted in the principles of SR but extends beyond traditional methodologies to address complex challenges in physics. It does so by leveraging a framework that iteratively refines feature selection, equation generation, and model optimization. This iterative process aims to construct a comprehensive model by aggregating simpler sub-models, each capturing unique aspects of the underlying physical processes.

\begin{algorithm}[h!]
\caption{MISR}

\KwIn{Dataset $D$ with features $X$ and target variable $Y$.}
\KwData{Hyperparameters: Number of folds $k$, number of terms in symbolic regression $n_t$, feature subset size $s$, maximum iterations $max_{iter}$, improvement ratio threshold $\theta$.}

\textbf{Feature Importance Assessment}:\\
\Indp Use Random Forest Regressor and Mutual Information regression on training data to evaluate feature importances.\\
\Indm \textbf{Iterative Model Building}:\\
\While{iteration $\leq max_{iter}$ and improvement ratio $\geq \theta$}{
    Perform k-fold cross-validation on the training set.\\
    \For{each fold}{
        Select a subset of features $X_{sub}$ of size $s$ based on importance scores.\\
        Conduct SR with $n_t$ terms on $X_{sub}$ to predict $Y$.\\
        Store top-performing equations based on the multi-objective evaluation.\\
    }
    Select the best-performing equation across all folds.\\
    Update $Y$ to be the residuals of the current best model.\\
    Increment iteration.\\
}
\textbf{Hyperparameter optimization}:\\
\Indp Utilize the validation dataset to fine-tune $k$, $n_t$, $s$, $max_{iter}$, and $\theta$.\\
\Indm \textbf{Model Finalization}:\\
\Indp Aggregate the equations from each iteration to form the final model.
\end{algorithm}

A key aspect of our methodology is the dynamic assessment of feature importance in each step. For this, we fit a Boosted Decision Tree's feature importances averaged with the Mutual Information score of each input feature~\cite{gregorutti2017correlation,peng2005feature}. 
Each iteration involves k-fold cross-validation, where we downsample the feature space based on its importance. 
\subsubsection*{SR framework}
Within each fold and iteration, we perform a symbolic regression constrained by the number of terms—a significant hyperparameter influencing equation simplicity. 
Each expression acts as an individual within the evolutionary algorithm, its fitness gauged via the mean error description length (MEDL) based on optimized parameters \cite{doiudrescu2020ai,udrescu2020ai}. These regressions are executed in parallel across folds for computational efficiency. The top-performing equations are identified based on their performance score on various objective variables which are related by our physical models, with experimental uncertainties inversely weighing the evaluation for robustness. 
\subsubsection*{Iterative refinement and residual modeling}
Our iterative approach refines the model's target variable to be the residuals from the best-performing model of the preceding iteration by repeating the fitting process in the residuals. The iterative process continues until reaching a predefined maximum number of steps or when the improvement ratio diminishes below a certain threshold.
\subsubsection*{Estimation of weighting distribution}
\begin{figure}[h!]
    \centering
    \includegraphics[width=0.5\textwidth]{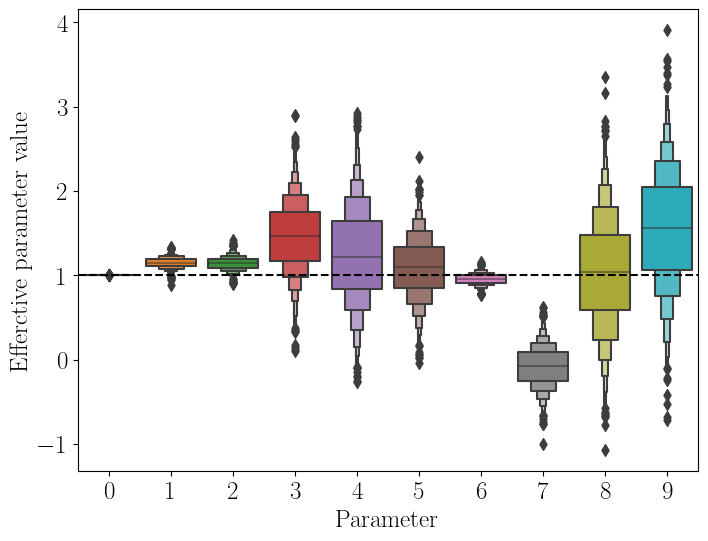}
    \caption{Distribution of sampled weights for charge radii's regressor after performing bootstrapping on Discriminative Jackknife. See the text for more details.}
    \label{fig:unc_weights}
\end{figure}

After we have modeled all of the expanding terms in our expression, we estimate the errors via the technique of Discriminative Jackknife~\cite{alaa2020discriminative} on the test set. Then the inference uncertainty is obtained via resampling from the distribution of the weights modeled as Gaussian distributions and adding the subsequent term value in the expansion cutoff. Fig.~\ref{fig:unc_weights} illustrates the distribution of sampled weights for charge radii's regressor. The figure demonstrates how the Jackknife resampling allows the assignment of characteristic Gaussian distributions to each one of the terms in the MISR expansion.

The total related uncertainty is computed by approximating all of the sampling weights to a Gaussian distribution and bootstrapping to one $\sigma$ since we assume the independence in each distribution. Then, we orthonormally add the uncertainties with the approximation coming from the cutoff in the expansion (i.e. the next expansion term).

\subsubsection*{Nucleon separation energies}
\begin{figure}[h!]
    \centering\includegraphics[width=0.5\textwidth]{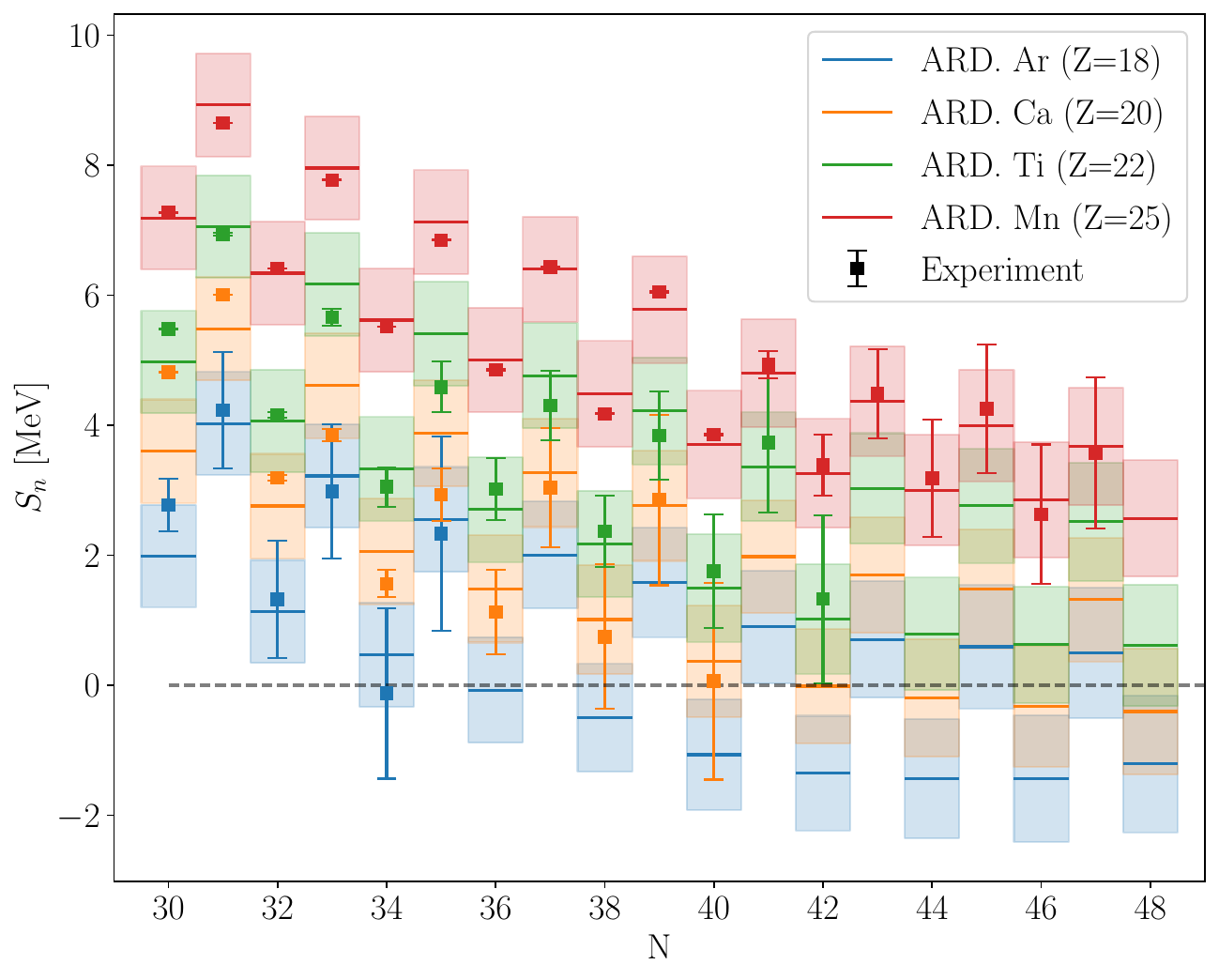}
    \caption{Experimental results for the neutron separation energy, $S_n$, are compared with the results from the ARD model for neutron-rich isotopes of Argonne (Ar), Calcium (Ca), Titanium (Ti), and Manganese (Mn). The uncertainties of the ARD model are shown as shaded areas.}
    \label{fig:sn_several}
\end{figure}

Fig.~\ref{fig:sn_several} presents the trend of neutron separation energies for different neutron-rich isotopes of Argonne (Ar), Calcium (Ca), Titanium (Ti), and Manganese (Mn). Experimental results are compared with the ARD model

\begin{figure}[h!]
    \centering\includegraphics[width=0.5\textwidth]{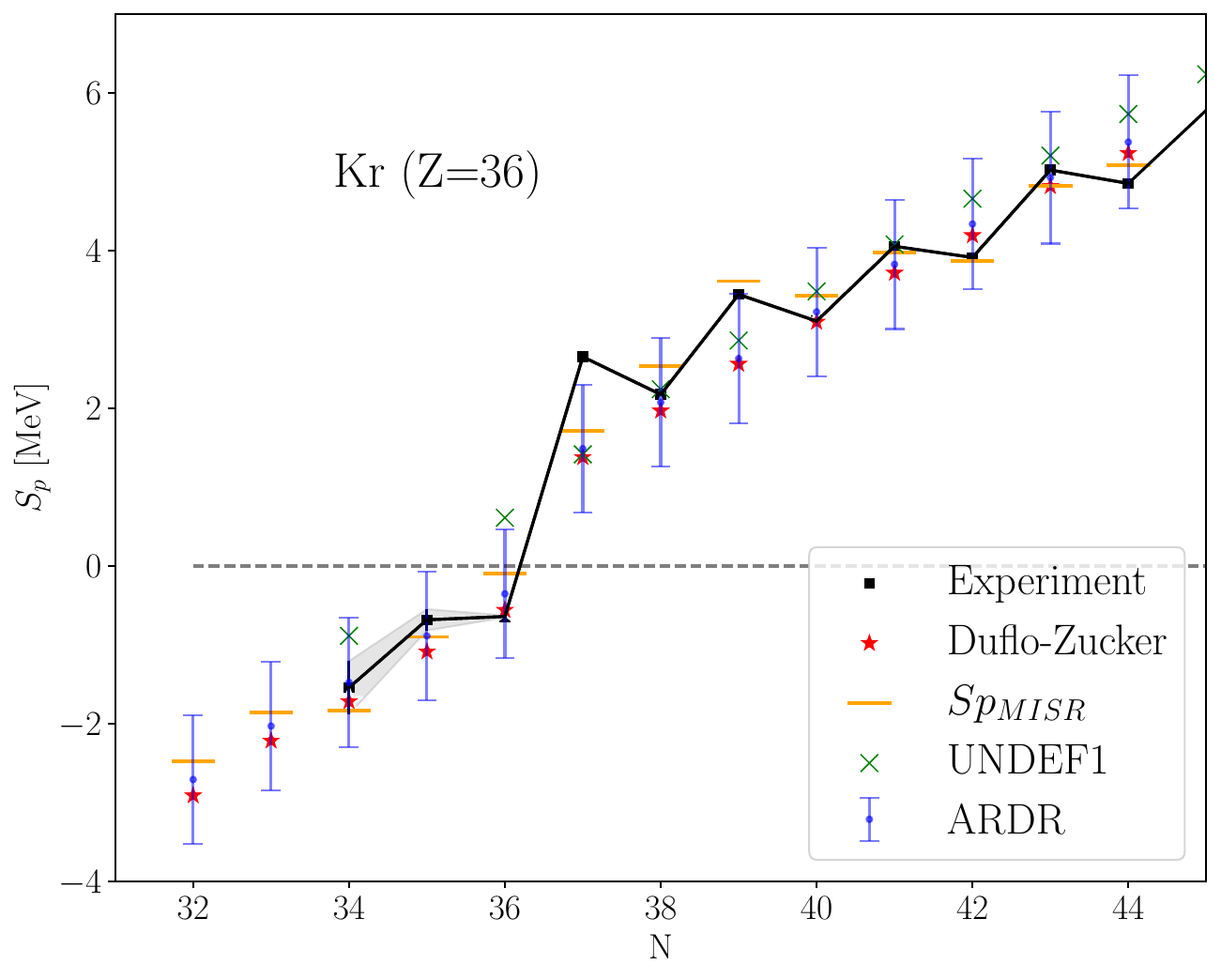}
    \caption{Proton separation energy of Krypton isotopes (Z=36). Experimental results are compared with our MISR and ARDR results. Results from the DZ model and DFT calculations are included for comparison.}
    \label{fig:sp_36}
\end{figure}

As a complementary example, the proton separation energies of Krypton isotopes are shown in Fig.~\ref{fig:sp_36}. 

\subsection{MISR for Binding Energy}

In table~\ref{tab:be_misr_eqs} we present the obtained $BE_{MISR}$ models term by term. We label the functions $\mod_2{(Z)}\equiv \gamma_Z$ and $\mod_2{(N)}\equiv\gamma_N$ respectively. In addition, we denote $\delta\equiv \gamma_Z-\gamma_N$ as is done traditionally in the literature.

\begin{table*}[h!]
    \centering
    \begin{tabular}{||c|c||}
        \hline\hline
         Model Order & Obtained Function \\
        \hline
        1 & $\left(\frac{Z}{N} + Z - \frac{1.06 N}{Z}\right) \left(I \left(32.4 - \frac{\sqrt[3]{A} N}{Z}\right) + 16.7\right)$    \\
        \hline
        2 & $3.42 (Z - 14.6) \left(\sqrt[3]{A} - 2.19 I - 4.38\right) (I - 0.110 \log (A)) + \delta - P + 0.301$\\
        \hline
        3 & $-2.02 e^{-0.40 \gamma_Z \gamma_N P - (0.040 Z)^Z} + 2.99 \cdot 0.867^{(N-Z)^2} - 0.426 P (\log (Z) - 3.30) + I$\\
        \hline
        4 & $A^{2/3} e^{-A^{2/3} + Z^{1.10} - Z} I \log \left(\frac{Z}{N}\right) + 0.634 e^{A^{2/3} + \sqrt[3]{A} - N} + 0.290 \gamma_N \gamma_Z + 0.246$\\
        \hline
        5 & $(0.0000154)^P A^{\frac{2 P}{3}} \left(P (N-Z)^2 + N\right) \left(0.0000154 (N-1)^2 + P\right)$\\
        \hline
        6 & $\frac{\gamma_N (1 - \gamma_Z)}{N} - \exp \left(\left(\sqrt[3]{A} - \frac{Z}{N} - 1.21\right) \left(2 (P + 0.108) \left(P^{\sqrt[3]{A}} - P^N\right) - \gamma_Z (1 - \gamma_N) - \frac{Z}{N} - 0.426\right)\right)$\\
        \hline
        7 & $1.35 I\left((0.324 - I) \left(-\frac{Z}{N} - 1.78\right) \left(\frac{4.30 N}{Z} - 0.111 \left(A + e^P\right)\right) - P + 1.35 I\right)$\\
        \hline
        8 & $(-0.801^{\gamma_N (1 - \gamma_Z)} + 0.570 P - 2 I) (-0.112 + (A - (N - Z)^2 + \frac{AN}{Z}) 0.801^N)$\\
        \hline
        9 & $9.20 \cdot 10^{-23} \cdot 1330^{A^{\frac{1}{3}}} (-1.97 + \gamma_N (-1 + \gamma_Z) - \gamma_Z + P) (-1330 + N^2 - 2NZ + Z^2) (-670 + N^2 - 2NZ + Z^2)$\\
        \hline
        10 & $3.02 \exp \left(-1.91 P^{1.94} \left(\frac{Z}{N}\right)^{A^{2/3}} - 0.895 e^{-0.227 N} N^2 - 0.0268 (N-1)^2\right)$\\
        \hline\hline
    \end{tabular}
    \caption{Analytical expression for the binding energy, $BE_{MISR}$, discovered by the MISR model for the first ten iterations. Each row represents a distinct iteration of the MISR process, showing the mathematical expression derived for each one. The expressions detail how the nuclear binding energy (BE) is modeled as a function of nuclear properties such as neutron number (N), proton number (Z), atomic mass ($A$), isospin asymmetry ($I$), and the Casten factor ($P$).}
    \label{tab:be_misr_eqs}
\end{table*}
\subsection{MISR for Charge Radii}

Table~\ref{tab:rc_misr_eqs} presents the obtained $rc_{MISR}$ models term by term. We label the functions $\mod_2{(Z)}\equiv \gamma_Z$ and $\mod_2{(N)}\equiv\gamma_N$ respectively. In addition, we denote $\delta\equiv \gamma_Z-\gamma_N$ as is done traditionally in the literature.

\begin{table*}[h!]
    \centering
    \begin{tabular}{||c|c||}
        \hline\hline
         Model Order & Obtained Function \\
        \hline
        1 & $A^{1/3} \left(0.950 + \frac{1.48}{Z}\right) + 0.0170 \left(P - \frac{Z}{N}\right) e^I - I$ \\
        \hline
        2 & $0.00400 \left(-\frac{\left(26.2 - P^2\right) \left(A + 26.2 \left(P^2 - \frac{2.35 N}{Z}\right)\right)}{Z^2} + \delta - 1.70\right)$ \\
        \hline
        3 & $0.0000762 (N_n - 1.81 N_p) (N_n + N_p - 3.26)$ \\
        \hline
        4 & $-0.00105 \left(\left(\frac{Z}{N}\right)^N - P + 1.61\right)$ \\
        \hline
        5 & $\frac{0.00200 P^2}{(N_p - 0.463) (P - 0.486)}$ \\
        \hline
        6 & $\frac{1}{\frac{19200 Z}{N} - 18100}$ \\
        \hline
        7 & $\frac{0.000588 N}{(A - 57.4) (N_n + 0.132)}$ \\
        \hline
        8 & $7860 \cdot 0.000118^N (\gamma_N(1 - \gamma_Z) + 0.508) \left(\frac{1}{N_n - 0.000220}\right)^N$ \\
        \hline
        9 & $1.64 (1 - \gamma_N)(1 - \gamma_Z) \left(-\frac{0.0195 Np}{Z} - 0.597\right)^N$ \\
        \hline
        10 & $\frac{N}{Z} \left(\frac{0.00520 N_n^{(\gamma_N (\gamma_Z - 1) - \gamma_Z + 2)}}{A} - 0.000400\right)$ \\
        \hline\hline
    \end{tabular}
    \caption{Analytical expression for the charge radius, $r_{C}$, discovered by the  MISR model for the first ten iterations.  It includes coefficients and terms involving atomic mass ($A$), proton number (Z), neutron number (N), the Casten factor ($P$), valence protons ($N_p$), and valence neutrons ($N_n$). The expressions are designed to incrementally capture the details of the nuclear charge radius, reflecting the refinement of the model through each iteration.}
    \label{tab:rc_misr_eqs}
\end{table*}

\end{document}